\documentstyle[a4,12pt,psfig]{article}

\newcommand{\gtrsim}{
\,\raisebox{0.35ex}{$>$}
\hspace{-1.7ex}\raisebox{-0.65ex}{$\sim$}\,
}

\newcommand{\lesssim}{
\,\raisebox{0.35ex}{$<$}
\hspace{-1.7ex}\raisebox{-0.65ex}{$\sim$}\,
}

\begin{document}

\bibliographystyle{prsty}



\title{ 
Magnetic free energy at elevated temperatures and hysteresis of magnetic particles
}      

\author{
H. Kachkachi$^*$ and D. A. Garanin$^{\dagger}$  
}
\date{}

\maketitle
$^*$Laboratoire de Magn\'{e}tisme et d'Optique, Univ. de Versailles St. Quentin, 
45 av. des Etats-Unis, 78035 Versailles, France 

$^{\dagger}$Max-Planck-Institut f\"ur Physik komplexer Systeme, N\"othnitzer Strasse 38,
D-01187 Dresden, Germany 

\begin{abstract}
We derive a free energy for weakly anisotropic ferromagnets which is valid in the
whole range of temperature and interpolates between the micromagnetic energy at zero temperature 
and the Landau free energy near the Curie point $T_c$.
This free energy takes into account the change of the magnetization length due to thermal
effects, in particular, in the inhomogeneous states.
As an illustration, we study the thermal effect on the Stoner-Wohlfarth curve and hysteresis 
loop of a ferromagnetic nanoparticle assuming that it is in a single-domain state.
Within this model, the saddle point of the particle's free energy, as well as the metastability
boundary, are due to the change in the magnetization length sufficiently close to $T_c$, as opposed to the usual homogeneous rotation process at lower temperatures. 
\end{abstract}     
\smallskip
\begin{flushleft}
PACS numbers: 75.10.-b, 75.50.Tt\\
Keywords: Thermal effects, hysteresis loop, Fine-particle systems
\end{flushleft}

\section{Introduction}
\label{introduction}

The macroscopic energy of weakly anisotropic ferromagnets which first appeared in the seminal
paper by Landau and Lifshitz \cite{lanlif35}, has been an instrument for
innumerable investigations of domain walls and other inhomogeneous states of magnetic systems.
Later an approach based on this macroscopic energy has been called
``micromagnetics'' \cite{bro63mic}.  
Strictly speaking, micromagnetics is an essentially zero-temperature theory for {\em
classical} magnets, as it considers the magnetization as a vector of fixed length.
Under these conditions, it can be easily obtained as a continuous limit of the classical
Hamiltonian on a lattice.
Practically, micromagnetics has been applied to nonzero temperatures as well, with
temperature-dependent equilibrium magnetization and anisotropy constants.

On the other hand, close to the Curie temperature $T_c$ a magnetic free energy of the
Landau type can be considered.
This free energy allows the magnetization to change both in direction and length, and it
is, in fact, a continuous limit of the free energy following from the mean-field
approximation (MFA).
Using this approach, Bulaevskii and Ginzburg \cite{bulgin63} predicted a phase transition between the
Bloch walls and the Ising-like walls at temperatures slightly below $T_c$. 
Much later this phase transition was observed experimentally
\cite{koegarharjah93,harkoegar95} using the theoretical results for the mobility of domain
walls in this regime \cite{gar91llb,gar91edw}.

Although both of these approaches have been formulated at the same mean-field level, they have
been considered as unrelated for a long time.
On the other hand, the MFA itself is an all-temperature approximation, and thus one can
question its macroscopic limit in the whole temperature range.
The corresponding macroscopic free energy should interpolate between the Landau-Lifshitz
energy, or micromagnetics at $T=0$ and the Landau free energy in the vicinity of $T_c$.
The Euler equation for the magnetization which minimizes this generalized free energy
appears in Refs.\ \cite{gar91llb} and \cite{gar97prb}.
Whereas the condition for the Landau theory is $M \ll M_s$, where $M_s$ is the saturation
magnetization at $T=0$, the new equations only require $|M-M_e| \ll M_s$, where $M_e$ is
the equilibrium magnetization at zero field.
For weakly anisotropic magnets (the anisotropy energy is much less than the homogeneous exchange
energy, and this is satisfied by most compounds) the magnetization magnitude $M$ is either
small or only slightly deviates from $M_e$, thus the condition above is satisfied in the
whole temperature range.

Derivation of the generalized macroscopic free energy from the MFA is much subtler
than that of the Euler equations.
This free energy appears in Ref.\ \cite{gar97prb} where its form was guessed.
In this paper, we will present this derivation and illustrate the resulting magnetic free
energy in the case of a single-domain magnetic particle with a uniaxial anisotropy.
For the sake of transparency, we will consider the exchange anisotropy rather than the single-site
anisotropy; this does not change the qualitative results.
It will be shown that the free-energy landscape for the magnetic particle has different
forms at low temperatures and near $T_c$.
At low temperatures, the saddle point between the ``up'' and ``down'' minima is located
near the sphere $|{\bf M}| = M_e$, the value of $M$ being somewhat reduced in comparison with
$M_e$ due to thermal effects at $T>0$.
Near $T_c$, the value of $M$ strongly changes by going from one minimum to another
through the saddle point; for zero fields the saddle point becomes $\bf M =0$. 
These effects also modify the metastability boundary of the magnetic particle (the famous
Stoner-Wohlfarth curve \cite{stowoh4891} which was experimentally observed in 
Ref.\ \cite{weretal97}), and its hysteresis curves.

The main body of this paper is organized as follows.
In Sec.\ \ref{secfenergy} we give the derivation of the magnetic free energy at all
temperatures.
In Sec.\ \ref{seclandscape} the free-energy landscape of a single-domain magnetic
particle is analyzed.
In Secs.\ \ref{secastroid} and \ref{sechysteresis} we study the Stoner-Wohlfarth curves and
hysteresis loops at different temperatures.
In Sec.\ \ref{HomoTherm} we discuss the possibility of observing the
new thermal effects for single-domain magnetic particles.

\section{Free energy of a magnetic particle}
\label{secfenergy}

Let us start with the {\em biaxial} ferromagnetic model described by the 
 classical anisotropic Hamiltonian of the type
\begin{eqnarray}\label{biaxham}
{\cal H} = - \mu_0{\bf H} \sum_i {\bf s}_i
- \frac{1}{2}\sum_{ij}J_{ij}
(s_{zi}s_{zj} + \eta_y s_{yi}s_{yj} + \eta_x s_{xi} s_{xj}) ,
\end{eqnarray}
where $\mu_0$ is the magnetic moment of the atom, $i,j$ are lattice sites, 
${\bf s}_i$ is the normalised vector, 
$|{\bf s}_i|=1$, and the dimensionless ani\-so\-t\-ro\-py factors satisfy 
$\eta_x \leq \eta_y \leq 1$.

To study the macroscopic properties of this system at nonzero
temperatures, it is convenient to use a macroscopic free energy.  
In the literature one can find two types of macroscopic free energies
for magnets.
One of them is the so-called micromagnetic free energy which is valid at zero
temperature, and the other one is Landau's free energy which is
applied near the Curie temperature $T_c$. 
Since both free energies are based on the mean-field approximation
(MFA), it is possible to derive a simple form of the MFA free
energy for weakly anisotropic ferromagnets which is valid in the whole
temperature range and bridges these two well-known forms.

The free energy $F=-T\ln {\cal Z}$ of a spin system 
described by the Hamiltonian in Eq.\ (\ref{biaxham}) 
can be calculated in the mean-field approximation by 
considering each spin on a site $i$ as an isolated 
spin in the effective field containing contributions determined 
by the mean values of the neighboring ones. Namely, 
\begin{equation}\label{hammfa}
{\cal H} \Rightarrow {\cal H}^{\rm MFA} 
= 
{\cal H}_{00} - \sum_i {\bf H}_i^{\rm MFA} {\bf s}_i ,
\end{equation}
where
\begin{equation}\label{ham00}
{\cal H}_{00} 
=
\frac{1}{2}\sum_{ij}J_{ij}
\left(
\sigma_{zi}\sigma_{zj} 
+ 
\eta_x \sigma_{xi} \sigma_{xj}
+ 
\eta_y \sigma_{yi}\sigma_{yj}
\right) , 
\end{equation}
$\mbox{\boldmath $\sigma$}_i \equiv \langle{\bf s}_i\rangle$ is the spin polarization, 
and the molecular field ${\bf H}_i^{\rm MFA}$ is given by
\begin{equation}\label{fieldmfa}
{\bf H}_i^{\rm MFA} = \mu_0 {\bf H} 
+ \sum_j J_{ij} 
\left(
\sigma_{zj} {\bf e}_z 
+ \eta_x \sigma_{xj} {\bf e}_x 
+ \eta_y \sigma_{yj} {\bf e}_y
\right) .
\end{equation}
Then the solution of the one-spin problem in Eq.\ (\ref{hammfa}) leads to
\begin{eqnarray}\label{fenergymicro}
&&
F = {\cal H}_{00} - NT\ln(4\pi) - T \sum_i \Lambda(\xi_i) \nonumber \\
&&
 \Lambda(\xi)\equiv \ln\left(\frac{\sinh(\xi)}{\xi}\right) ,
\end{eqnarray}
where $N$ is the total number of spins,
$\xi_i \equiv |\mbox{\boldmath $\xi$}_i|$, and
$\mbox{\boldmath $\xi$}_i \equiv \beta {\bf H}_i^{\rm MFA}$.
The MFA free energy determined by Eq.\ (\ref{fieldmfa}) and 
Eq.\ (\ref{fenergymicro}) can be minimized with respect to the spin 
averages $\mbox{\boldmath $\sigma$}_i$ to find the equilibrium solution 
in the general case where the anisotropy $1-\eta_{x,y}$ is not 
necessarily small.
The minimum condition for the free energy, 
$\partial F/\partial\mbox{\boldmath $\sigma$}_i=0$,
leads to the Curie-Weiss equation
\begin{equation}\label{cweissgen}
\mbox{\boldmath $\sigma$}_i 
=
B(\xi_i)\frac{\mbox{\boldmath $\xi$}_i}{\xi_i},
\end{equation}
where $B(\xi)=\coth(\xi) - 1/\xi$ is the Langevin function.

For small anisotropy, $1-\eta_{x,y} \ll 1$, one can go over to the 
continuum limit and write for the short-range interaction $J_{ij}$
\begin{equation}\label{spinlapl}
\sum_j J_{ij}\mbox{\boldmath $\sigma$}_j 
\cong 
J_0\mbox{\boldmath $\sigma$}_i 
+ 
J_0\alpha\Delta\mbox{\boldmath $\sigma$}_i ,
\end{equation}
where $\Delta$ is the Laplace operator acting on the components
of $\mbox{\boldmath $\sigma$}({\bf r})$, 
$J_0$ is the zero Fourier component (the zeroth moment), and 
$J_0\alpha$ is the second moment of the exchange interaction $J_{ij}$.
For the simple cubic lattice with nearest neighbor interactions $\alpha=a_0^2/z$,
$z=6$, and $a_0$ is the lattice spacing.
Going from summation to integration in Eq.\ (\ref{fenergymicro}), one obtains
\begin{eqnarray}\label{fenergymacro1}
\frac{F}{J_0} = - \frac{NT}{J_0}\ln(4\pi) + \frac{1}{v_0} \!\!\int\!\! d{\bf r}
\left\{
\frac{1}{2}\sigma^2 
+ 
\frac{1}{2}( \mbox{\boldmath $\sigma$},{\bf h}_{\rm eff}-{\bf h} )
- 
\frac{1}{\beta J_0} \Lambda(\xi)
\right\} ,
\end{eqnarray}
where $v_0$ is the unit-cell volume,
%
%
\begin{eqnarray}\label{defheff}
&&
\mbox{\boldmath $\xi$} 
= \beta J_0 ( \mbox{\boldmath $\sigma$} + {\bf h}_{\rm eff} )
\nonumber \\
&&
{\bf h}_{\rm eff} = {\bf h} + \alpha \Delta \mbox{\boldmath $\sigma$} 
- (1-\eta_x) \sigma_x {\bf e}_x - (1-\eta_y) \sigma_y {\bf e}_y ,
\end{eqnarray}
and ${\bf h} \equiv \mu_0 {\bf H}/J_0$.
We will consider the case of small fields, $h \ll 1$.
Since in this case in Eq.\ (\ref{defheff}) 
$|{\bf h}_{\rm eff}| \ll |\mbox{\boldmath $\sigma$}|$ in the whole 
range below $T_c$ (near $T_c$ the value of $\sigma$ is small but the susceptibility is
large), the last term of Eq.\ (\ref{fenergymacro1}) can be 
expanded to first order in ${\bf h}_{\rm eff}$ using 
%
%
\begin{equation}\label{deltaxi}
\xi = \xi_0 + \delta\xi, \qquad
\xi_0 = \beta J_0 \sigma, \qquad
\delta\xi 
\cong 
\beta J_0 \frac{\mbox{\boldmath $\sigma$}{\bf h}_{\rm eff}}{\sigma}
\end{equation}
and the first two terms of the expansion
%
%
\begin{equation}\label{devlam}
\Lambda(\xi) 
\cong 
\Lambda(\xi_0) 
+ 
B(\xi_0)\delta\xi 
+ 
\frac{1}{2}B'(\xi_0)(\delta\xi)^2 ,
\end{equation}
where $B'(\xi)\equiv dB(\xi)/d\xi$.
Hence, 
%
%
\begin{eqnarray}\label{fenergymacro}
\frac{F}{J_0} &=& \frac{1}{v_0} \!\!\int\!\! d{\bf r}
\left\{\frac{1}{2}\sigma^2 -\frac{1}{\beta J_0} \Lambda(\xi_0) -
\frac{B(\xi_0)}{\sigma}\mbox{\boldmath $\sigma$}{\bf h} \right.
\left.-
\left(\frac{B(\xi_0)}{\sigma}-\frac{1}{2}\right)
( \mbox{\boldmath $\sigma$},{\bf h}_{\rm eff}-{\bf h} )
\right\}\nonumber \\ 
&-& \frac{NT}{J_0}\ln(4\pi). 
\end{eqnarray}
Near $T_c = T_c^{\rm MFA}= J_0/3$  the order parameter $\mbox{\boldmath $\sigma$}$ becomes small, 
and using $\Lambda(\xi) \cong \xi^2/6 - \xi^4/180$ and
$B(\xi_0) \cong \xi_0/3 \cong \mbox{$\beta_c J_0\sigma/3$} = \sigma$ 
one straightforwardly arrives at the Landau free energy
%
%
\begin{eqnarray}\label{fenergylan}
\frac{F}{J_0} &=& \frac{1}{v_0} \!\!\int\!\! d{\bf r} \left\{
\frac{1}{2}\alpha (\nabla \mbox{\boldmath $\sigma$})^2 - 
\mbox{\boldmath $\sigma$}{\bf h} + 
\frac{1}{2}(1-\eta_x) \sigma_x^2 \right.
\left.+ \frac{1}{2}(1-\eta_y) \sigma_y^2
- 
\frac{\epsilon}{2} \sigma^2 + \frac{3}{20} \sigma^4
\right\} \nonumber \\
&-& \frac{NT}{J_0}\ln(4\pi) ,
\end{eqnarray}
where $\epsilon \equiv (T_c^{\rm MFA}-T)/T_c^{\rm MFA}$ and 
$({\bf \nabla \sigma})^{2}=({\bf \nabla }\sigma_{x})^{2}+({\bf \nabla }
\sigma_{y})^{2}+({\bf \nabla }\sigma_{z})^{2}$.
Note that Eq.\ (\ref{fenergylan}) formally yields unlimitedly increasing values of $\sigma$ at equilibrium,
as a function of the field $h$.
To comply with the condition $\sigma \ll 1$ in Landau's formalism, $h$ should be kept
small, as was required above.
One could also work out the $h^2$ corrections to  Eq.\ (\ref{fenergylan}).

Now we consider the temperature region where the influence of anisotropy 
and field on the magnitude of the spin polarization $\sigma$ can be studied perturbatively.
Concerning the influence of anisotropy, the applicability criterion can 
be obtained from the requirement that the local shift of $T_c$ for spins 
forced perpendicularly to the easy axis should be smaller than the 
distance from $T_c$, i.e., $\Delta T_c/T_c \sim 1-\eta \ll \epsilon$. 
The macroscopic free energy in the perturbative region can then be combined with 
the Landau free energy in their common applicability range $1-\eta \ll \epsilon \ll 1$. 
Thus, in the perturbative region we expand the first two terms of expression 
Eq.\ (\ref{fenergymacro}) up to the second order in $\delta\sigma\equiv\sigma-\sigma_e$ using
%
%
\begin{equation}\label{deltaxie}
\xi_0 = \xi_e + \delta\xi, \qquad
\xi_e = \beta J_0 \sigma_e, \qquad
\delta\xi \cong \beta J_0 \delta\sigma 
\end{equation}
and the formula analogous to Eq.\ (\ref{devlam}).
This leads to
%
%
\begin{eqnarray}\label{fenergymacro2}
\frac{F}{J_0} = \frac{F_e}{J_0} + \frac{1}{v_0} \!\!\int\!\! d{\bf r}
\left\{
\frac{1}{2}(1-B'\beta J_0)(\sigma-\sigma_e)^2 \right.
\left.-\frac{B(\xi_0)}{\sigma}\mbox{\boldmath $\sigma$}{\bf h}
-\left(\frac{B(\xi_0)}{\sigma}-\frac{1}{2}\right)
( \mbox{\boldmath $\sigma$},{\bf h}_{\rm eff}-{\bf h} )
\right\} ,
\end{eqnarray}
where $B'=B'(\xi_e)$, and
%
%
\begin{equation}\label{feequi}
\frac{F_e}{J_0} = - \frac{NT}{J_0}\ln(4\pi) 
+ N
\left[
\frac{1}{2}\sigma_e^2 - \frac{1}{\beta J_0}\Lambda(\xi_e) 
\right], 
\end{equation}
is the equilibrium free energy in the absence of magnetic field and
the quantity $\sigma_e$ is the spin polarisation at equilibrium satisfying the homogeneous
Curie-Weiss equation
%
%
\begin{equation}\label{curieweisseq}
\sigma_e=B(\xi_e).
\end{equation}
The minimum condition for Eq.\ (\ref{fenergymacro2}),
$\delta F/\delta \mbox{\boldmath $\sigma$}=0$,
after an accurate calculation taking into account the dependence of 
$B(\xi_0)$ on $\sigma$ and neglecting  the terms quadratic in 
${\bf h}_{\rm eff}$, results in an equation of the form
%
%
\begin{equation}\label{diffeqmag}
\frac{1}{\bar\chi_\|}
(\sigma-\sigma_e)\frac{\mbox{\boldmath $\sigma$}}{\sigma}
- 
{\bf h}_{\rm eff}
+
\frac{
[\mbox{\boldmath $\sigma$}
\times [\mbox{\boldmath $\sigma$}
\times {\bf h}_{\rm eff}]]
}{\sigma^2\bar\chi_\|} = 0 ,
\end{equation}
where ${\bf h}_{\rm eff}$ is given by Eq.\ (\ref{defheff})
and the dimensionless longitudinal susceptibility $\bar\chi_\|$ is given in Eq.\ (\ref{barchi}) below.
The solution of Eq.\ (\ref{diffeqmag}) satisfies 
$\mbox{\boldmath $\sigma$}\| {\bf h}_{\rm eff}$,
and the term with the double vector product plays no role.
Considering the response to small fields ${\bf h}= h_z {\bf e}_z$ and 
${\bf h}= h_{x,y} {\bf e}_{x,y}$
in Eq.\ (\ref{diffeqmag}) in a homogeneous situation
(in the transverse case ${\bf h}_{\rm eff}=0$ and $\sigma=\sigma_e$), 
one can identify the reduced susceptibilities for the spin polarization as
%
%
\begin{eqnarray}\label{barchi}
&&
\bar\chi_\| \equiv \frac{d\sigma_z}{dh_z} = \frac{B'\beta J_0}{1-B'\beta J_0},
\nonumber \\
&&
\bar\chi_x = \frac{1}{1-\eta_x},
\qquad
\bar\chi_y = \frac{1}{1-\eta_y} .
\end{eqnarray}

Our expression for the free energy, Eq.\ (\ref{fenergymacro2}), is still
cumbersome, but can be simplified if we make the observation 
that in the perturbative region the deviation 
$\delta\sigma\equiv \sigma-\sigma_e$ 
is proportional to $h_{\rm eff}$, and
the terms of the type 
$\delta\sigma \cdot h_{\rm eff}$ and $(\delta\sigma)^2$ 
in Eq.\ (\ref{fenergymacro2}) are thus quadratic in $h_{\rm eff}$.
Such terms are nonessential in the calculation of $F$ itself, they are 
only needed for the proper writing of the equilibrium equation 
Eq.\ (\ref{diffeqmag}).
Now we can replace Eq.\ (\ref{fenergymacro2}) by a simplified form 
%
%
\begin{eqnarray}\label{fenergymacro3}
\frac{F}{J_0} = \frac{F_e}{J_0}
+ \frac{1}{v_0} \!\!\int\!\! d{\bf r}
\left\{
\frac{1}{2}\alpha (\nabla \mbox{\boldmath $\sigma$})^2 
- 
\mbox{\boldmath $\sigma$}{\bf h} + 
\frac{1}{2}(1-\eta_x) \sigma_x^2 \right.
\left.+\frac{1}{2}(1-\eta_y) \sigma_y^2 
+ 
\frac{1}{2\bar\chi_\|}(\sigma-\sigma_e)^2
\right\} .
\end{eqnarray}
This form coincides with Eq.\ (\ref{fenergymacro2}), if we set 
$\sigma=\sigma_e$, i.e., 
it yields the same value of $F$ in the leading first order in 
$h_{\rm eff}$.
On the other hand, Eq.\ (\ref{fenergymacro3}) leads to the same equilibrium 
equation, Eq.\ (\ref{diffeqmag}), without the nonessential last term.
Now, at the last step of the derivation, one can combine 
Eq.\ (\ref{fenergymacro3}) with the Landau free energy in Eq.\ (\ref{fenergylan}), 
leading to
%
%
\begin{eqnarray}\label{fenergymacro4}
\frac{F}{J_0} = \frac{F_e}{J_0}
+ \frac{1}{v_0} \!\!\int\!\! d{\bf r}
\left\{
\frac{1}{2}\alpha (\nabla \mbox{\boldmath $\sigma$})^2 
- 
\mbox{\boldmath $\sigma$}{\bf h} 
+ \frac{1}{2\bar\chi_x} \sigma_x^2 \right.
\left.+ \frac{1}{2\bar\chi_y} \sigma_y^2 
+ \frac{1}{8\sigma_e^2\bar\chi_\|}(\sigma^2-\sigma_e^2)^2
\right\} .
\end{eqnarray}
Indeed, in the Landau region, $\epsilon \ll 1$, from 
Eq.\ (\ref{curieweisseq}) and Eq.\ (\ref{barchi}) it follows $\sigma_e^2\cong (5/3)\epsilon$ 
and $\bar\chi_\|\cong (2\epsilon)^{-1}$, and Eq.\ (\ref{fenergymacro4}) simplifies 
to Eq.\ (\ref{fenergylan}). 
In terms of the magnetization ${\bf M}$ defined by 
\begin{equation}\label{defM}
{\bf M}({\bf r}) = \mu_0 \mbox{\boldmath $\sigma$}({\bf r})/v_0,
\qquad  \mbox{\boldmath $\sigma$}({\bf r}) \equiv \langle {\bf s}_{\bf
r}\rangle,
\end{equation}
and other dimensional quantities, the free energy Eq.\ (\ref{fenergymacro4}) takes on the form 
\cite{harkoegar95,gar97prb}
\begin{eqnarray}\label{fenergymacro5}
F = F_e + \!\!\int\!\! d{\bf r}
\left\{
\frac{1}{2 q_d^2}(\nabla {\bf M})^2 - {\bf M} \mbox{\boldmath $\cdot$} {\bf H} 
+ \frac{1}{2\chi_x} M_x^2 \right.
\left.+ \frac{1}{2\chi_y} M_y^2 
+ \frac{1}{8 M_e^2\chi_\|}(M^2-M_e^2)^2
\right\},
\end{eqnarray}
with
%
%
\begin{equation}\label{fenergyident}
q_d^2 = \frac{W_D}{\alpha J_0},
\qquad
\chi_\alpha = \frac{W_D}{J_0}\bar\chi_\alpha,
\qquad
W_D \equiv \frac{(\mu_0)^2}{v_0},
\end{equation}
where $q_d$ is the so-called dipolar wave number, $W_D$ is the characteristic energy
of the dipole-dipole interaction, and $\chi_\alpha \equiv dM_\alpha/dH_\alpha$ with
$\alpha=x,y,z$ are the susceptibilities [cf. Eq.\ (\ref{barchi})].
The applicability of Eq.\ (\ref{fenergymacro5}) requires that the deviation
$M-M_e$ from the equilibrium magnetization $M_e$ in the absence of anisotropy and field, 
is small in comparison with the saturation value $M_s=\mu_0/v_0$.
This is satisfied in the whole range of temperature if the anisotropy and field are small, i.e.,  
$1-\eta_{x,y} \ll 1$ and $\mu_0H \ll J_0$.
On the other hand, the free energy in Eq.\ (\ref{fenergymacro5}) can be transformed into the ``micromagnetic'' 
form by introducing the magnetization direction vector
$\mbox{\boldmath$\nu$} \equiv {\bf M}/M$.
One can then write
%
%
\begin{equation}\label{identmicromag}
\frac{1}{2\chi_{x,y}} M_{x,y}^2 = K_{x,y} \nu_{x,y}^2,
\qquad
K_{x,y} = \frac{M^2}{2\chi_{x,y}},
\end{equation}
where $K_{x,y}$ are the anisotropy constants.
In particular, for the uniaxial model one can rewrite
%
%
\begin{equation}\label{identuniax}
\frac{1}{2\chi_\perp} (M_x^2 + M_y^2) = - K \nu_z^2 + K,
\end{equation}
where $K$ is the uniaxial anisotropy constant.
Note that at nonzero temperatures $M$ and thus the anisotropy constants can be spatially
inhomogeneous.
In this case Eq.\ (\ref{fenergymacro5}) is more useful than its micromagnetic form.

It should be stressed that the traditional way of writing the magnetic free energy in
terms of the magnetization is, at least from the theoretical point of view, somewhat
artificial.
All terms in Eq.\ (\ref{fenergymacro5}) apart from the Zeeman term, are non-magnetic
and in fact independent of the atomic magnetic moment $\mu_0$.
The latter cancels out of the resulting formulae, as soon as they are reexpressed in terms
of the original Hamiltonian (\ref{biaxham}).
On the other hand, Eq.\ (\ref{fenergymacro5}) is convenient if the parameters are taken
from experiments.

Examination of the formalism above shows that it can be easily generalized to
quantum systems, leading to the same form as in Eq.\ (\ref{fenergymacro5}). 
In the derivation, the classical Langevin function $B(x)$ is replaced by the quantum
Brillouin function $B_S(x)$.

\section{The free-energy landscape for a uniaxial magnetic particle}
\label{seclandscape}
 
Henceforth, we will consider the uniaxial anisotropy, $\eta_x=\eta_y\equiv
\eta_\perp$.
For single-domain magnetic particles in a homogeneous state, the gradient terms in the free energy
can be dropped and the free energy of Eq.\ (\ref{fenergymacro5}) can be
presented in the form  
\begin{eqnarray}\label{freduced}
&&
 F = F_e + (VM_e^2/\chi_\perp) f \nonumber \\
&&
f=-{{\bf n} \cdot {\bf
h}}+\frac{1}{2}(n_{x}^{2}+n_{y}^{2})+\frac{1}{4a}(n^{2}-1)^{2},
\end{eqnarray}
where $V$ is the particle's volume and $f$ the reduced free energy written in
terms of the reduced variables
%
%
\begin{equation}\label{defnha}
{\bf n} \equiv {\bf M}/M_e, 
\qquad {\bf h} \equiv {\bf H}\chi_\perp/M_e,
\qquad a \equiv 2\chi_\|/\chi_\perp.
\end{equation}
One can see that the parameter $a$ here controls the rigidity of the magnetization vector; 
it goes to zero in the zero-temperature limit (the fixed magnetization length) and
diverges at $T_c$ as $a\cong (1-\eta)/\epsilon$ within the MFA.
As the MFA is not quantitatively accurate, it is better to consider the susceptibilities
and hence $a$ as taken from experiments.
Although this procedure is not rigorously justified, it can improve the results.
Note that the reduced free energy $f$ in Eq.\ (\ref{freduced}) is only defined
for $T<T_c$ since the reduced magnetization ${\bf n}$ is normalized by $M_e$.

For fields ${\bf h}$ inside the Stoner-Wohlfarth astroid, which will be
generalized here to nonzero temperatures, $f$ has two minima separated by a barrier.
Owing to the axial symmetry, one can set $n_y=0$ for the investigation of the free 
energy landscape.
The minima, saddle points, and the maximum can be found from the equations  
$\partial f/\partial n_{x}=\partial f/\partial n_{z}=0$, or, explicitly
\begin{eqnarray}\label{eqextrema}
&&
n_{z}(n^2-1) = ah_z \nonumber \\ 
&&
n_{x}(n^2-1+a)=ah_x.
\end{eqnarray}
These equations can be rewritten as 
%
%
\begin{equation}\label{eqextrema2}
1-n^2 = -ah_z/n_z = a(1 - h_x/n_x),
\end{equation}
whereupon the important relation follows
%
%
\begin{equation}\label{xzrel}
\frac{h_x}{n_x} - \frac{h_z}{n_z} = 1 \qquad {\rm or} \qquad
n_x = \frac{h_x n_z}{h_z + n_z}.
\end{equation}
Using the latter, one can solve for $n_x$ and obtain a 5th-order equation for
$n_z$
%
%
\begin{equation}\label{nzeq}
h_x^2 n_z^3 = (h_z+n_z)^2(ah_z + n_z-n_z^3),
\end{equation}
from which the parameters of the potential landscape such as energy minima, saddle
point, and energy barriers can be found.

In zero field, the characteristic points of the energy landscape can be simply
found from Eqs.\ (\ref{eqextrema}).
One of these points is $n_x=n_z=0$, which is a local maximum for $a<1$ and a saddle point
for $a>1$. 
The minima are given by $n_x=0$, $n_z=\pm 1$.
The saddle points correspond to $n_z=0$, while from the second of Eqs.\
(\ref{eqextrema}) one finds
\begin{eqnarray}\label{nxsaddle}
n_x = 
\left\{
     \begin{array}{ll}
            \pm\sqrt{1-a},  &  a \leq 1 \\
            0,              &  a \geq 1.
     \end{array}
\right.
\end{eqnarray}
In fact, due to the axial symmetry, for $a<1$ one has  a saddle circle
$n_x^2+n_y^2=1-a$ rather than two saddle points.
The free-energy barrier following from this solution is given by
%
%
\begin{equation}\label{febarrier}
\Delta f \equiv f_{\rm sad} - f_{\rm min} = 
\left\{
\begin{array}{ll}
(2-a)/4,          &   a \leq 1 \\
1/(4a),           &  a \geq 1.
\end{array}
\right.
\end{equation}
\begin{figure}[t]
\unitlength1cm
\begin{picture}(11,6.5)
\centerline{\psfig{file=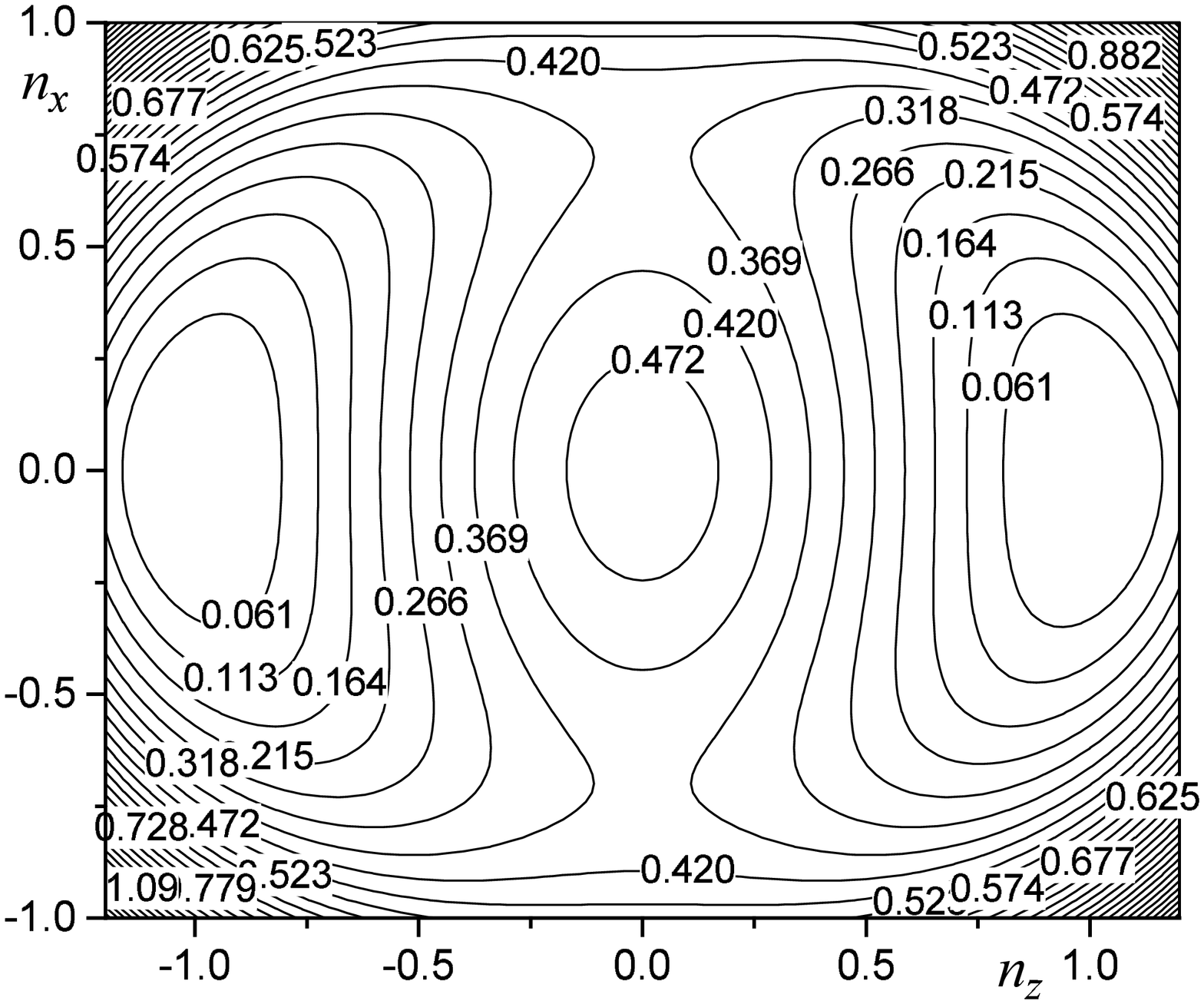,angle=-90,width=8cm}}
\end{picture}
\begin{picture}(11,6)
\centerline{\psfig{file=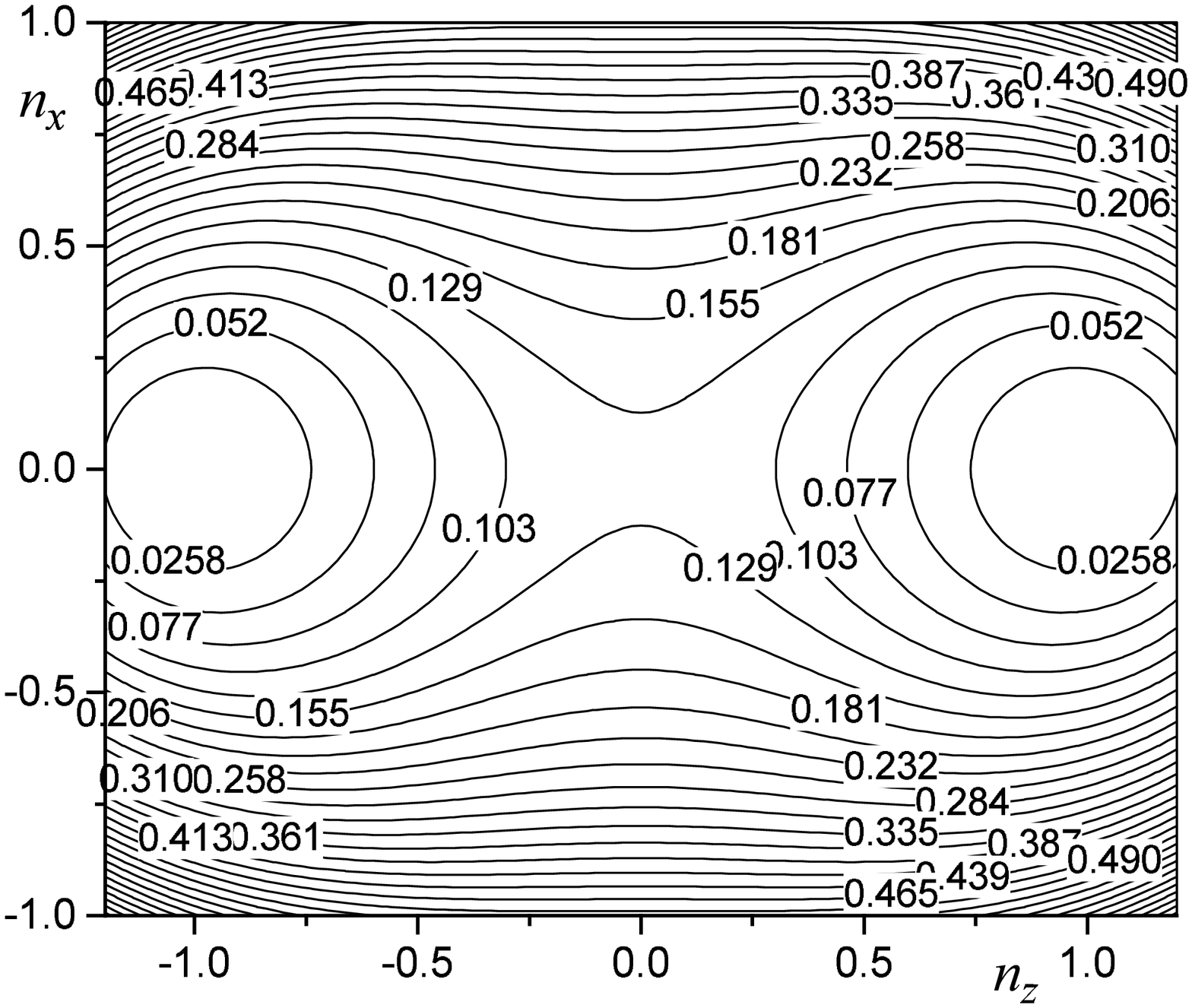,angle=-90,width=8cm}}
\end{picture}
\caption{ \label{tsw_land}
The free energy of a ferromagnetic particle with uniaxial anisotropy
[$f$ in Eq.\ (\protect\ref{freduced}) in zero field] for $a\equiv 2\chi_\|/\chi_\perp = 0.5$ (upper plot)
and $a=2$ (lower plot) corresponding to lower and higher temperatures, respectively.
}
\end{figure}

The free-energy landscape in zero field is shown in Fig.\ \ref{tsw_land}.
At nonzero temperatures $a>0$, the magnitude of the magnetization at the saddle is smaller than
unity since it is directed perpendicularly to the easy axis, and for this
orientation the ``equilibrium" magnetization is smaller than in the direction
along the $z$ axis.
For $a>1$, the two saddle points, or rather the saddle circle, degenerate into
a single saddle point at $n_x=n_z=0$, and the local maximum there disappears.
That is, for the magnetization to overcome the barrier, it is easier to change
its magnitude than its direction.
This is a phenomenon of the same kind as the phase transition in ferromagnets between the
Ising-like domain walls in the vicinity of $T_c$ (the magnetization changes its
magnitude and is everywhere directed along the $z$ axis) and the Bloch walls at
lower temperatures \cite{bulgin63,koegarharjah93}.

\section{The Stoner-Wohlfarth curve}
\label{secastroid}

The Stoner-Wohlfarth curve separates the regions where there are two minima and one
minimum of the free energy.
On this curve the metastable minimum merges with the saddle point and loses its
local stability. The corresponding condition is
%
%
\begin{equation}\label{swcond}
\partial ^{2}f/\partial n_{x}^{2}\times \partial ^{2}f/\partial n_{z}^{2}
-(\partial ^{2}f/\partial n_{x}\partial n_{z})^{2} = 0,
\end{equation}
or, explicitly 
%
%
\begin{equation}\label{swcond2}
(1-n^{2})\left( 1-3n^{2}-a\right) +2an_{z}^{2}=0.
\end{equation}
Using Eq.\ (\ref{eqextrema2}), one can transform the equation above to the
quartic equation for $n_z$
%
%
\begin{equation}\label{swcondnz}
h_z[(2+a)n_z + 3ah_z] + 2n_z^4=0.
\end{equation}
Before considering the general case, let us analyze the limiting cases $a\ll 1$
and $a\gg 1$.

At low temperatures, i.e., $a\ll 1$, the magnetization only slightly deviates from its
equilibrium value, and from Eq.\ (\ref{swcond2}), to first order in $a$, one obtains
%
%
\begin{equation}\label{nrel}
n^2 \cong 1 - an_z^2 \qquad {\rm or} \qquad
n_x^2 + (1+a) n_z^2 \cong 1.
\end{equation}
From Eq.\ (\ref{swcondnz}) and the analogous equation for $n_x$ and
using Eq.\ (\ref{eqextrema2}), one can derive
the field dependence of $n_z$ and $n_x$ on the Stoner-Wohlfarth curve
\begin{eqnarray}\label{nznxonsw}
&&
n_z \cong -h_z^{1/3} [1 - (a/2)(h_z^{2/3}-1/3)] \nonumber \\
&&
n_x \cong h_x^{1/3} [1 + (a/2)(h_x^{2/3}-1)].
\end{eqnarray}
Inserting these results in Eq.\ (\ref{nrel}), one arrives at the equation for the 
Stoner-Wohlfarth astroid
%
%
\begin{equation}\label{swasrtsmall}
h_x^{2/3} + [(1+a/2)h_z]^{2/3}  \cong 1, \qquad a\ll 1,
\end{equation}
where $a$ is given in Eq.\ (\ref{defnha}).
One can see that, in comparison with the standard zero-temperature Stoner-Wohlfarth astroid,
i.e. at $a=0$, $h_z$ is rescaled.
The critical field in the $z$ direction decreases because of the field dependence of the 
magnetization magnitude at nonzero temperatures.

In the case $a\gg 1$, i.e. near $T_c$, Eq.\ (\ref{swcond2}) relating $n_x$ and
$n_z$ in the equation for Stoner-Wohlfarth curve simplifies to
%
%
\begin{equation}\label{nrel1}
n_x^2 + 3 n_z^2 \cong 1.
\end{equation}
Using this equation together with Eqs.\ (\ref{eqextrema2}), one obtains
%
%
\begin{equation}\label{nznxonsw1}
n_z \cong -(ah_z/2)^{1/3}, \qquad n_x \cong h_x.
\end{equation}
Making use of this result in Eq.\ (\ref{nrel1}), one obtains another limiting case
of the Stoner-Wohlfarth curve
%
%
\begin{equation}\label{swalarge}
h_x^2 + 3(ah_z/2)^{2/3} \cong 1, \qquad a\gg 1.
\end{equation}
In this case, the critical field (i.e., the field on the Stoner-Wohlfarth curve) 
in the $z$ direction is strongly reduced, and
there is no singularity in the dependence $h_{cz}(h_x)$ at $h_x=0$.

\begin{figure}[t]
\unitlength1cm
\begin{picture}(11,6)
\centerline{\psfig{file=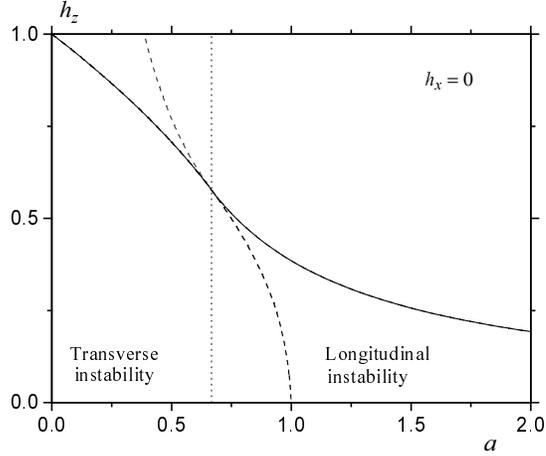,angle=-90,width=8cm}}
\end{picture}
\caption{ \label{tsw_hz}
Dependence $h_z(a)$ at $h_x=0$ on the Stoner-Wohlfarth curve.
}
\end{figure}
\begin{figure}[t]
\unitlength1cm
\begin{picture}(11,7)
\centerline{\psfig{file=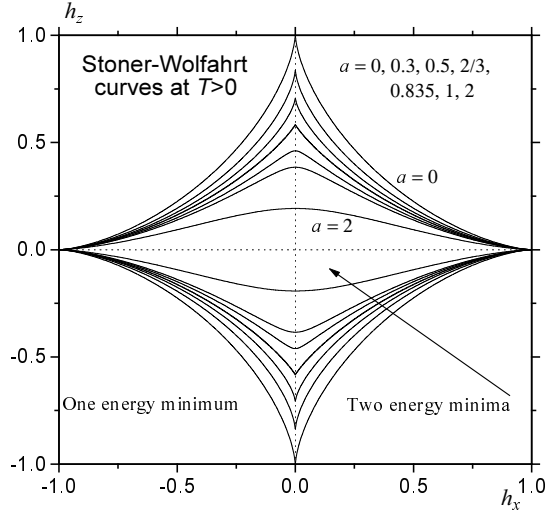,angle=-90,width=8cm}}
\end{picture}
\caption{ \label{tsw_tsw}
The Stoner-Wohlfarth curves at different temperatures, $a\equiv 2\chi_\|/\chi_\perp=0$
($T=0$), 0.3, 0.5, 2/3, 0.835, 1, 2.
}
\end{figure}

The qualitatively different character of the Stoner-Wohlfarth curves in these two
cases is due to the different mechanisms pertaining to the loss of the local stability for
the field applied along the $z$ axis.
For $h_x=0$ the mixed derivative $(\partial ^{2}f/\partial n_{x}\partial n_{z})$
in  Eq.\ (\ref{swcond}) vanishes, and Eq.\ (\ref{swcond}) factorizes.
Explicitly, we have
%
%
\begin{equation}\label{swcondfact}
(a-1+n_z^2)(-1+3n_z^2)=0.
\end{equation}
Vanishing of the first factor in this equation corresponds to the loss of
stability with respect to the rotation of the magnetization, 
$\partial ^{2}f/\partial n_{x}^{2}=0$.
Vanishing of the second factor, $\partial ^{2}f/\partial n_{z}^{2}=0$, 
implies the loss of stability with respect to the change of the magnetization length.
Using $n_x=0$, with the help of the first equality in Eqs.\ (\ref{eqextrema2}) one
obtains in the two cases
%
%
\begin{equation}\label{hzc}
h_{cz} = 
\left\{
\begin{array}{ll}
h_{z\|} \equiv \sqrt{1-a},              & a \leq 2/3 \\
h_{z\perp} \equiv 2/(3^{3/2}a),      & a \geq 2/3.
\end{array}
\right.
\end{equation}
Note that the transition between the two regimes occurs here at a different
value of $a$ than in Eq.\ (\ref{nxsaddle}). 
The dependence $h_z(a)$ at $h_x=0$ is shown in Fig.\ \ref{tsw_hz}.

In the general case, it is easier to find the Sto\-ner-\-Wohlfarth curve numerically
from Eqs.\ (\ref{eqextrema}) and (\ref{swcond2}).
The results in the whole range of $a$ are shown in Fig.\ \ref{tsw_tsw}.

\begin{figure}[t]
\unitlength1cm
\begin{picture}(11,6)
\centerline{\psfig{file=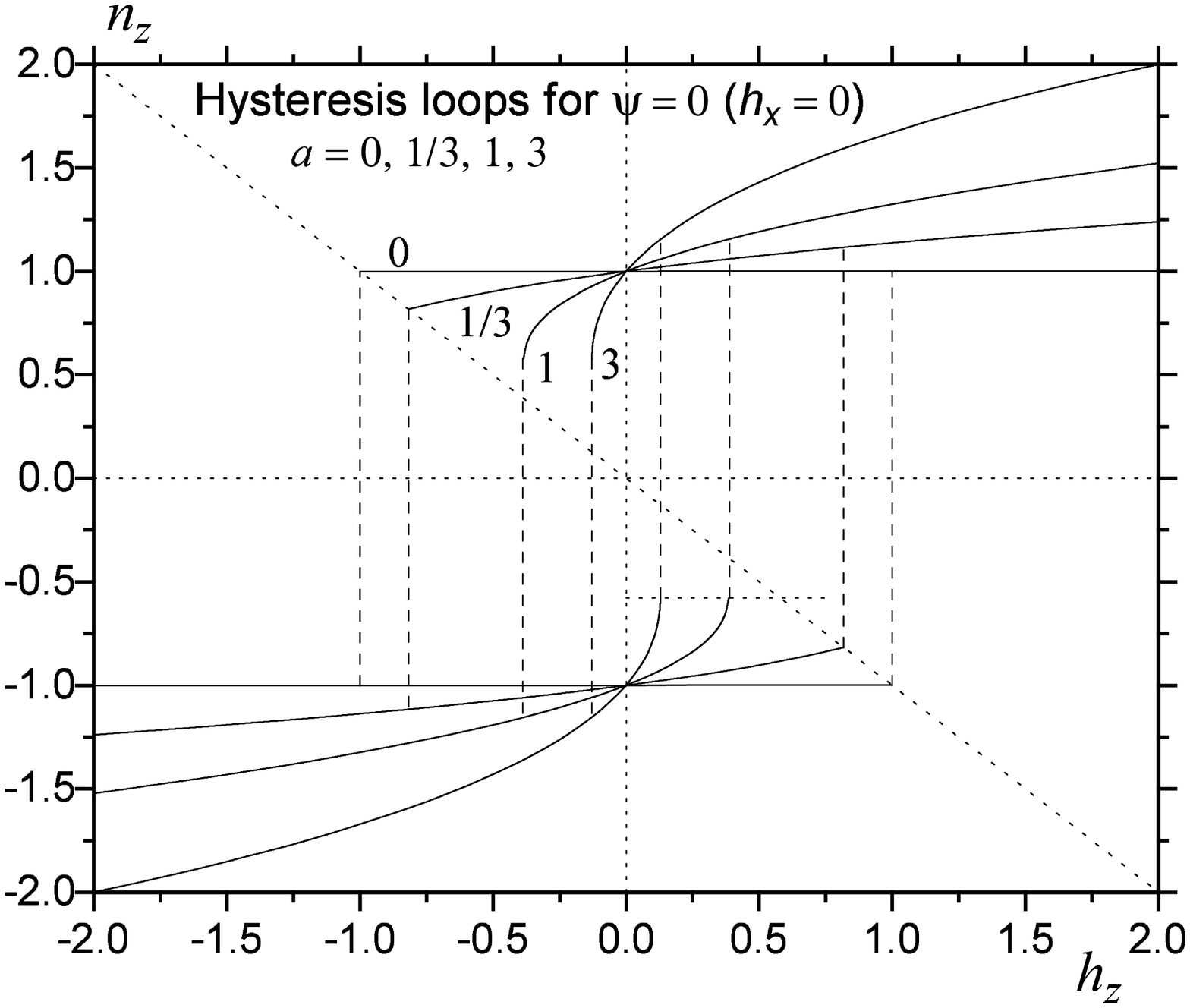,angle=-90,width=8cm}}
\end{picture}
\begin{picture}(11,6)
\centerline{\psfig{file=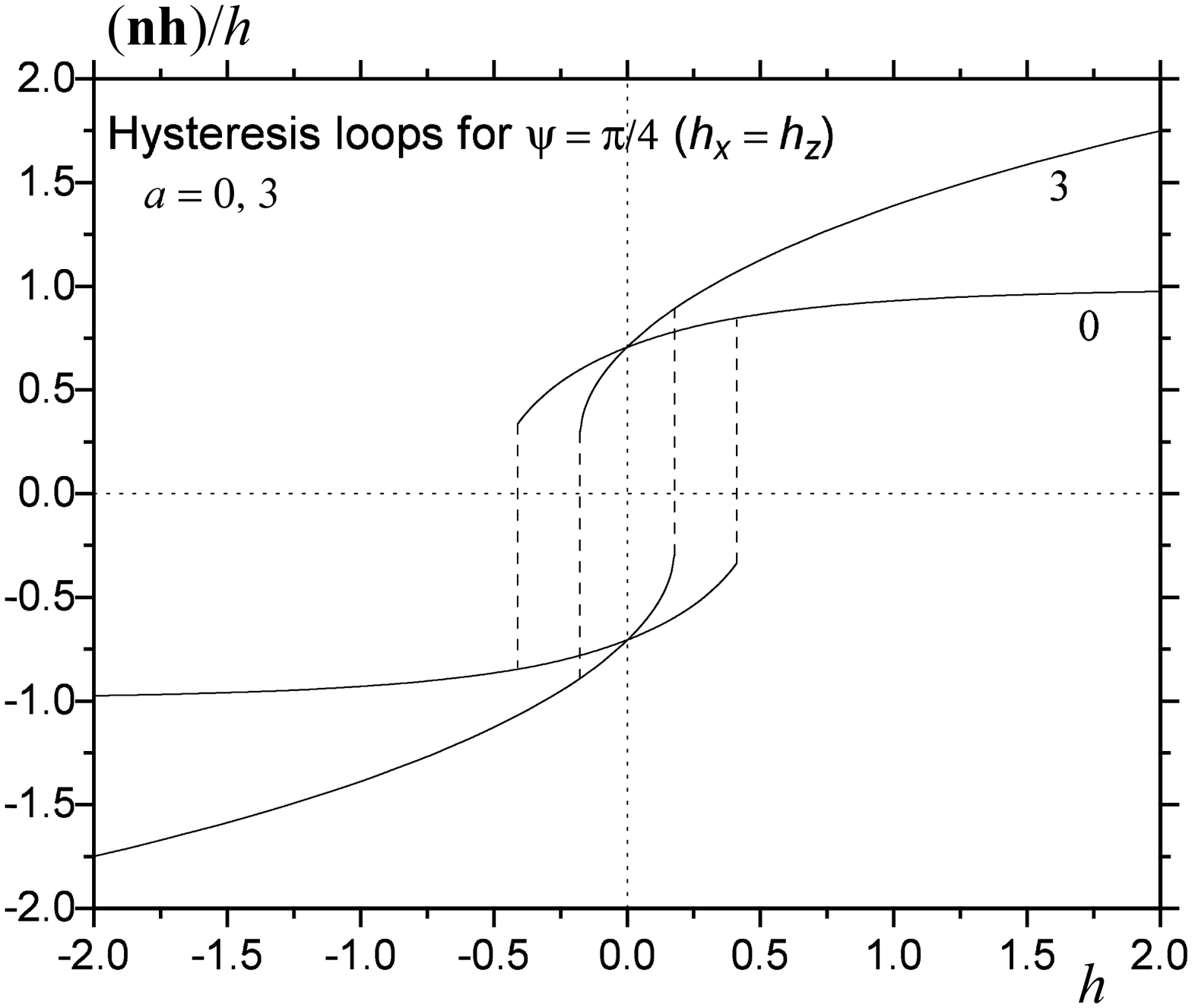,angle=-90,width=8cm}}
\end{picture}
\caption{ \label{tsw_hy}
Thermal effect on the static hysteresis loops [$a(0)=0$, $a(T_c)=\infty$]. 
Upper plot: $\psi=0$ (${\bf h}\| {\bf e}_z$), lower plot: $\psi=\pi/4$.
}
\end{figure}

\section{Hysteresis loops}
\label{sechysteresis}

In this section we use our model to study the hysteresis loop, i.e., the dependence 
${\bf n(h)}$ at the stable or metastable free-energy minimum.
At first we consider the case $h_x=0$ in which the problem can be solved analytically.
Setting $n_x=0$ in the first of Eqs.\ (\ref{eqextrema}) one obtains the cubic equation
$n_z^3-n_z-ah_z=0$, the solution of which, for the positive branch of the hysteresis curve,
reads
%
%
\begin{equation}\label{hyst}
n_z = 
\left\{
\begin{array}{ll}
\displaystyle
\frac{2}{\sqrt{3}} \cos \left(\frac{\phi}{3}\right),   & \left| h_z\right| \leq h_{z\|}     \\
\displaystyle
\left[\frac{ah_z}{2}+\sqrt{D}\right]^{1/3} +
\left[\frac{ah_z}{2}-\sqrt{D}\right]^{1/3},  & \left| h_z\right|  \geq h_{z\|},
\end{array}
\right.
\end{equation}
where $h_{z\|}$ is given by Eq.\ (\ref{hzc}) and 
%
%
\begin{equation}\label{defDphi}
\phi \equiv \arccos(h_z/h_{z\|}), \qquad D \equiv (a/2)^2 (h_z^2 - h_{z\|}^2).
\end{equation}
The negative branch of the hysteresis curve can be obtained by the reflection
$h_z\Rightarrow -h_z$ and $n_z\Rightarrow -n_z$.
Eq.\ (\ref{hyst}) is written in the form which is explicitly real.
In fact, both forms hold in the whole range $h_z \geq -h_{z\|}$, and there is no change of
behavior at $h_z = h_{z\|}$, as expected on physical grounds.
The characteristic values of the magnetization on the positive branch are 
$n_z(-h_{z\|}) = 1/\sqrt{3}$ and $n_z(h_{z\|})= 2/\sqrt{3}$.
The solution above was obtained by setting $n_x=0$ and thus ignoring the transverse
instability which can occur before the longitudinal instability at $h_z = -h_{z\|}$ for
the positive branch.
This competition of instabilities has been studied in the previous section.
It was found that for $a<2/3$ the transverse instability occurs at the fields
$|h_z|= |h_{z\|}|=\sqrt{1-a} < |h_{z\perp}|$.
Thus in this case the branches of the hysteresis curves should be cut at $\pm h_{z\|}$;
at these fields the system jumps to the other branch.

The analytical results obtained above for $h_x=0$ and the numerical ones for the case of ${\bf h}$
directed at the angle $\psi = \pi/4$ to the easy axis are shown in Fig.\ \ref{tsw_hy}.
For $h_x=0$, the derivative $dn_z/dh_z$ diverges at the metastability
boundary for $a\geq 2/3$ (longitudinal instability).

\section{The homogeneity criterion, thermal activation, spin-waves}
\label{HomoTherm}

In the preceding sections we illustrated how the general magnetic free energy of 
Eq.\ (\ref{fenergymacro5}) works for the simplest model of a uniaxial magnetic particle
in a single-domain state at elevated temperatures.
Here we discuss the possibility of observing the new types of behavior found above.
This requires satisfying two rather restrictive conditions.
First, the particle spends in the metastable minimum a time that is long enough for
measuring only if the barrier height energy is much larger than thermal energy.
At any $T>0$, the particle will escape from the metastable state via
thermal activation with a rate exponentially small at low temperatures.
For this reason, strictly speaking, the static hysteresis at nonzero temperatures does not
exist and dynamic measurements are needed.
At elevated temperatures, the required frequency of these measurements can become too
large.   
Second, the free energy barrier in the single-domain state should be smaller than the
barrier energy related with the formation of a domain wall which would travel through the particle and
switch the magnetization from one state to the other. 
The two criteria can be combined as follows
%
%
\begin{equation}\label{Criteria}
T \ll \Delta F_{SD} < \Delta F_{DW}.
\end{equation}
The first criterion here requires that the particle's volume is high enough whereas the
second criterion requires that it does not exceed some maximal value.
Let us consider, for instance, the zero field case, in which $F_{SD}$ is the 
single-domain free-energy barrier of Eqs.\ (\ref{febarrier}) and (\ref{freduced}). 
At not too high temperatures $a \lesssim 1$, one has $\Delta f \sim 1$, and the domain
walls are usual Bloch walls with the energy per unit area
%
%
\begin{equation}\label{DWEnergy}
w = 2M_e^2/(q_d^2 \delta), \qquad \delta = \sqrt{\chi_\perp}/q_d,
\end{equation}
where $\delta$ is the domain-wall width. 
For a spherical particle of radius $R$, the saddle point of the energy
corresponds to the domain wall through the center of the particle.
Using the definitions of parameters introduced in Sec.\ \ref{secfenergy}, one obtains the
formula 
%
%
\begin{equation}\label{EnergyRatio}
\frac{ \Delta F_{SD} }{ \Delta F_{DW} } = \frac{ Rq_d }{ 12 \chi_\perp^{1/2} } 
\sim \sqrt{1-\eta} \frac{ R }{ a_0 },
\end{equation}
where $a_0$ is the lattice spacing.
It is seen that the single-domain behavior of particles with $R \gg a_0$ requires small
values of the anisotropy $1-\eta$.
Working out the first inequality in Eq.\ (\ref{Criteria}), one can rewrite these equations in the
form
%
%
\begin{equation}\label{Criteria1}
\left( \frac{ \theta }{ (1-\eta) \sigma_e^2 } \right)^{1/3}  
\ll \frac{ R }{ a_0 } \lesssim \frac 1 {\sqrt{1-\eta}} ,
\end{equation}
where $\theta \equiv T/T_c^{\rm MFA}$ and $\sigma_e$ is the spin polarization at
equilibrium.
Clearly, for $\theta \ll 1$ these conditions can be satisfied.
Observing the new effects exhibited by the hysteresis suggested above requires 
$a \equiv 2\chi_\|/\chi_\perp  \equiv 2\bar\chi_\|/\bar\chi_\perp \sim 1$, which for
small anisotropy, i.e. $1-\eta \ll 1$, requires approaching $T_c$. 
Indeed, near $T_c$ the first of Eqs.\ (\ref{barchi}) yields
$\bar\chi_\| \cong (2\epsilon)^{-1}$, thus $a \sim 1$ implying $\epsilon \sim 1-\eta$.
In this region in Eq.\ (\ref{Criteria1}) one has $\theta \sim 1$ and
$\sigma_e^2\cong (5/3)\epsilon$. 
Then the existence of the interval for $R/a_0$ in Eq.\ (\ref{Criteria1}) requires 
$(1-\eta)/\epsilon \ll \epsilon$ which is impossible since in this region 
$(1-\eta)/\epsilon \sim 1$ and $\epsilon \ll 1$.
A similar analysis shows that also in the region where $a > 1$ Eqs.\ (\ref{Criteria})
cannot be satisfied.

Thus we are led to the conclusion that the qualitatively different types
of behavior of single-domain magnetic particles for $a\gtrsim 1$ cannot be observed with the standard
techniques.
If the particle's size is small enough, the single-domain criterion is satisfied, but
increasing temperature to $a\sim 1$ causes strong thermal fluctuations.
The potential landscape shown in Fig.\ \ref{tsw_land} is still valid but for studying the dynamics
of the magnetic particle in this range we need a special kind of Fokker-Planck
equation for {\em non-rigid} magnetic moments. Such an equation has not been considered yet.

On the other hand, for large particle sizes the energy barriers are high and thus the process
of thermal activation is suppressed, which is favorable for the observation of hysteresis
loops.
In this case, however, the barrier states are those with a domain wall across the
particle.
Analyzing these states goes beyond the scope of this article.
We only mention that for $a> 1/2$ the structure of domain walls in a ferromagnet is
completely different from that of a Bloch wall: The transverse magnetization component in
the wall is zero everywhere whereas the longitudinal component changes its magnitude and goes through
zero in the center of the wall \cite{bulgin63}.
For the observation of thermal effects in the hysteresis via inhomogeneous states, higher
values of the anisotropy $1-\eta$ are needed.
In this case, thermal effects manifest themselves starting from low temperatures.
An example of a strongly anisotropic material is Co, for which one obtains $1-\eta \simeq 0.02$.
This value is still much smaller than one, so that the validity condition for the magnetic
free energy of Eq.\ (\ref{fenergymacro5}) is satisfied.

Returning to the results of this paper, we can say that one can only observe corrections
to the well-known results, such as that of Eq.\ (\ref{swasrtsmall}), in the range 
$a \ll 1$, i.e., not close to $T_c$.  
On the other hand, one should not forget that the mean-field approximation used in this paper, 
while leading to qualitatively correct predictions, fails to account for some subtler
effects that 
can be responsible for important modifications of the results.
One of these effects is the influence of spin waves on the longitudinal susceptibility
which enters the definition $a \equiv 2\chi_\|/\chi_\perp$. 
Whereas within the MFA $\chi_\|$ rapidly decreases with decreasing temperature below
$T_c$ and is independent of the anisotropy, it is {\em infinite} in the whole range below $T_c$ for 
isotropic ferromagnets because of spin waves.
The square-root singularity in the dependence $M(H)$ at zero field has been
experimentally observed and reported on in Ref.\ \cite{koegoedompie94}.
In uniaxial ferromagnets, $\chi_\|$ becomes large for small anisotropies. 
This means that if the mean-field expression for $a$ is replaced by its value taken from
experiments, which is much larger due to spin-wave effects, the thermal effects
discussed in this paper will considerably increase in intensity.
Such a redefinition of $a$ is, of course, not rigorously justified, although it captures
the essential physics.
A more involved approach taking into account spin-wave effects for an exactly solvable model confirms the
concomitant increase of thermal effects on the variation of the magnetization length, as
was explicitly shown for domain walls in Ref.\ \cite{gar96jpa}.

Finally, we would like to mention that quite recently, experimental
results have been obtained by Wernsdorfer 
et al. \cite{WW00} on 3 nm cobalt nanoparticles which clearly show the
disappearance of the singularity near $H_x = 0$ at a temperature circa
8 K (the blocking temperature being 14 K). The height of the
experimental astroid decreases nearly as its width with increasing
temperature, but it does not become flat as predicted by our
calculations, which is not surprising considering the fact that $T\ll T_c$, but the
disappearance of the singularity is definitive.

\section{Conclusion}
\label{secconclusion}

In this paper we have derived a macroscopic free energy for
weakly anisotropic ferromagnets which is based on the mean-field approximation and is valid
in the whole range of temperature interpolating between the micromagnetic energy at
$T=0$ and the Landau free energy near $T_c$.
As an illustration, we have considered single-domain magnetic particles with uniaxial
anisotropy and we have shown that thermal effects qualitatively change the free-energy
landscape at sufficiently high temperatures, so that the passage from one free-energy minimum to the
other is realized by the {\em uniform change of the magnetization length} rather than the {\em uniform
rotation}. 
This also qualitatively changes the character of the Stoner-Wohlfarth curve and hysteresis
loops.
The latter effects cannot be observed with standard methods, however, because keeping the height of the
free-energy barrier much larger than thermal energy requires so large particle sizes that the
single-domain criterion is no longer satisfied.
For the uniform states, the theory is valid at low temperatures, but then the thermal effects
considered in the paper are small corrections to the zero-temperature results.
Large thermal effects on hysteresis should be searched for at temperatures close to $T_c$
in particles
of larger sizes, where the saddle point of the free energy is an {\em inhomogeneous}
state.
Investigation of the corresponding more complicated processes is beyond the scope of this
paper.

\section*{Acknowledgements}
D. A. Garanin is indebted to Laboratoire de Magn\'etisme et d'Optique
for the warm hospitality extended to him during his stay in Versailles in January 2000.


\begin{thebibliography}{10}

\bibitem{lanlif35}
{L. D. Landau and E. M. Lifshitz}, Phys. Z. Sowjetunion {\bf 8},  153  (1935).

\bibitem{bro63mic}
{W. F. Brown, Jr.}, {\em Micromagnetics} (Interscience, New York, 1963).

\bibitem{bulgin63}
{L. N. Bulaevskii and V. L. Ginzburg}, Zh. Eksp. Teor. Fiz. {\bf 45},  772
  (1963) [JETP {\bf 18},  530  (1964)].

\bibitem{koegarharjah93}
{J. K\"otzler, D. A. Garanin, M. Hartl, and L. Jahn}, Phys. Rev. Lett. {\bf
  71},  177  (1993).

\bibitem{harkoegar95}
{M. Hartl-Malang, J. K\"otzler, and D. A. Garanin}, Phys. Rev. B {\bf 51},
  8974  (1995).

\bibitem{gar91llb}
{D. A. Garanin}, Physica A {\bf 172},  470  (1991).

\bibitem{gar91edw}
{D. A. Garanin}, Physica A {\bf 178},  467  (1991).

\bibitem{gar97prb}
{D. A. Garanin}, Phys. Rev. B {\bf 55},  3050  (1997).

\bibitem{stowoh4891}
{E. C. Stoner and E. P. Wohlfarth}, Philos. Trans. R. Soc. London, Ser. A {\bf
  240},  599  (1948);
 IEEE Trans. Magn. {\bf MAG-27},  3475
  (1991).

\bibitem{weretal97}
{W. Wernsdorfer, E. Bonet Orozco, K. Hasselbach, A. Benoit, B. Barbara, N.
  Demoncy, A. Loiseau, and D. Mailly}, Phys. Rev. Lett. {\bf 78},  1791
  (1997).

\bibitem{koegoedompie94}
{J. K\"otzler, D. G\"orlitz, R. Dombrowski, and M. Pieper}, Z. Phys. B {\bf
  94},  9  (1994).

\bibitem{gar96jpa}
{D. A. Garanin}, J. Phys. A {\bf 29},  2349  (1996).

\bibitem{WW00}
{W. Wernsdorfer et al.}, preprint January 2000.

\end{thebibliography}

Electronic addresses: \\
$^*$kachkach@physique.uvsq.fr\\
$^{\dagger}$garanin@mpipks-dresden.mpg.de; http://www.mpipks-dresden.mpg.de/$\sim$garanin/

\end{document}